# Detecting Privileged Documents by Ranking Connected Network Entities

Jianping Zhang
Lega Technology and Data Analytics
Ankura Consulting Group, LLC
Washington, D.C., USA
jianping.zhang@ankura.com

Han Qin
Lega Technology and Data Analytics
Ankura Consulting Group, LLC
Washington, D.C., USA
han.qin@ankura.com

Nathaniel Huber-Fliflet
Lega Technology and Data Analytics
Ankura Consulting Group, LLC
London, UK
nathaniel.huber-fliflet@ankura.com

*Abstract*—**This paper presents a link analysis approach for identifying privileged documents by constructing a network of human entities derived from email header metadata. Entities are classified as either counsel or non-counsel based on a predefined list of known legal professionals. The core assumption is that individuals with frequent interactions with lawyers are more likely to participate in privileged communications. To quantify this likelihood, an algorithm assigns a score to each entity within the network. By utilizing both entity scores and the strength of their connections, the method enhances the identification of privileged documents. Experimental results demonstrate the algorithm's effectiveness in ranking legal entities for privileged document detection.**

*Keywords – privileged document detection, link analysis, eDiscovery, attorney work product, privilege, link analysis*

## I. Introduction

In the United States, companies responding to litigation or a request from an enforcement agency, typically are obligated to produce to the requesting party all nonprivileged material relevant and proportional to the legal case. To fulfill this obligation, corporate legal teams are typically responsible for identifying, collecting, and reviewing substantial volumes of business records to determine which documents are relevant to the legal case.

When attorneys are reviewing to identify relevant documents, they also typically work to identify which documents contain privileged material. In most cases, any document containing privileged material can and should be withheld from production. Privileged materials are also generally 'protected' from disclosure, by the attorney-client privilege and the work product doctrine. With data volumes increasing in legal matters, privilege review places a large burden on legal counsel when performing the document review – further increasing cost.

In today's legal environment, privileged materials are most often emails and electronic documents that either involve communications with attorneys or were created at their request in connection with actual or anticipated legal matters. Because many privileged documents consist of lawyer-related communications, one effective method for automatically identifying them is to detect exchanges between individuals associated with legal roles.

This paper introduces a link analysis method for identifying privileged documents by constructing a network of human entities based on email header metadata. Using a list of known lawyers, each entity in the network is classified as either counsel or non-counsel. Our method generates the lawyer list by extracting names from email signatures using an advanced legal signature detection algorithm. The underlying assumption is that individuals with frequent connections to lawyers are more likely to be involved in privileged communications. An algorithm was developed to assign a score to each entity in the human entity network, reflecting the likelihood of their involvement in creating or communicating in privileged documents—the higher the score, the greater the probability. A score for each link between entities can be derived from the scores of the two connected entities. By leveraging both entity and link scores, we can more effectively identify documents that are likely to be privileged.

In Section II, we provide a brief overview of the existing approaches to ranking and classifying nodes within connected networks. We then introduce our algorithm for scoring and ranking human entities based on a network constructed from email header metadata extracted from documents. Section IV presents the results of our experiments, which were conducted using data from three real-world cases to evaluate the proposed method. Finally, Section V offers our conclusions and outlines directions for future research.

## II. Ranking and Classification of Nodes in Network Analysis

PageRank [1][2] is one of the most popular algorithms for ranking network nodes. PageRank is a foundational algorithm developed by Larry Page and Sergey Brin, the co-founders of Google, to rank webpages in search results. It operates on the principle that the importance of a webpage can be inferred from the number and quality of links pointing to it. By treating the web as a directed graph, where pages are nodes and hyperlinks are edges, PageRank assigns a numerical score to each page based on its connectivity. The algorithm iteratively calculates a score for each page, with a page's score being influenced by the number and rank of the pages that link to it. The algorithm models a "random surfer" who navigates the web by clicking links at random, with occasional jumps to arbitrary pages, capturing both popularity and relevance. This probabilistic approach revolutionized search technology by prioritizing authoritative content and remains influential in network analysis and information retrieval today.

Another popular work in link analysis was introduced in [3] which introduces algorithms for identifying authoritative web pages by analyzing the web's link structure. It establishes a mutually reinforcing relationship: authoritative pages are rich in content on a given topic, while hub pages serve as directories that link to multiple authorities. In this framework, a strong authority is one that is linked to by many high-quality hubs, and a strong hub is one that links to many reputable authorities.

Another piece of related work [4] is collective classification. Collective classification is a machine learning approach used to predict the labels of multiple interconnected instances (such as nodes in a network) simultaneously, rather than independently. It leverages not only the features of each individual instance but also the relationships and labels of neighboring instances to improve prediction accuracy.

## III. Ranking Legal Entities

This section presents a straightforward algorithm for scoring legal entities within a network constructed from a collection of documents, including email communications. The legal entity network is derived by analyzing metadata in documents and the relationships embedded in these sources. Each node in the network represents a person, and a directed link between two nodes corresponds to an email sent from one individual to another. All links are directed.

Each node in the network is labeled as either "counsel" or "non-counsel." Nodes identified as counsel are assigned an initial score of 1, while non-counsel nodes begin with a score of 0. Email communications between two counsel-designated entities are more likely to be privileged than those between non-counsel entities. We further assume that communications between entities with higher scores are more likely to be privileged than those involving entities with lower scores.

The proposed legal entity ranking algorithm assumes that an entity's score can be inferred from the scores of its connected entities. If an entity is linked to numerous counsel-designated nodes, it is more likely to be involved in privileged communications and should therefore receive a higher score. The proposed algorithm is conceptually similar to Google's PageRank. PageRank works on the principle that a webpage is considered important if it is linked to by many other important pages. Table 1 shows the legal entity ranking algorithm.

In this algorithm, a node's new score is computed as 30% of its previous score plus 70% of the average score of all its connected nodes. High-scoring nodes are expected to be connected to many other high-scoring nodes. The use of max(10, N.nConnected) ensures that a node receives a high score only if it is connected to at least 10 other high-scoring nodes. The algorithm iteratively updates each node's score until the number of iterations reaches the user-defined parameter, MaxIteration. Initially, only counsel nodes are assigned non-zero scores. After the first iteration, nodes connected to counsel nodes begin to accumulate non-zero scores. With each subsequent iteration, more nodes receive non-zero scores as the influence propagates through the network.

TABLE I. Legal Entity Ranking Algorithm

```
Initialize all counsel node scores as 1
Initialize all non-counsel node scores as 0
Repeat
    Iteration = 0
    For each node N:
        N.NewScore = 0
        N.nConnected = 0
        For each link L of N:
            L.NodeFrom.NewScore = L.NodeFrom.NewScore +
                                        L.NodeTo.NewScore
            L.NodeTo.NewScore = L.NodeTo.NewScore +
                                        L.NodeFrom.NewScore
            L.NodeFrom.nConnected = L.NodeFrom.nConnected + 1
            L.NodeTo.nConnected = L.NodeTo.nConnected + 1
    Iteration = iteration + 1
    For each node N:
        N.score = 0.3*N.score + 0.7*N.NewScore/max(10,
                                        N.nConnected)
Until MaxIteration == Iteration
```

## IV. Experiments

We conducted experiments to evaluate the legal entity ranking algorithm using three datasets derived from separate real-world matters. The first dataset contains 104,753 documents, approximately 28% of which are privileged. The second dataset includes 395,167 documents, with around 16.5% classified as privileged. The third dataset consists of 100,727 documents, among which 24.4% are privileged. The majority of these documents are email communications. A network of interconnected entities is constructed from the email metadata in each dataset, where a directed link is established between two entities whenever one sends an email to the other.

### A. First Experiment

The first set of experiments examines the correlations between entity scores and both the number of privileged documents, and the precision associated with each entity. Similarly, it explores the relationship between link scores and the number of privileged documents, as well as link-level precision. Each link score is calculated as the average of the scores of the two connected entities.

Figures 1 & 2 illustrate the correlations between entity scores and both the number of privileged documents and precision for the first dataset. These metrics—score, document count, and precision—were calculated as averages across groups of 1,000 entities. The figures display the curves corresponding to one, two, and three iterations of the algorithm. As scores increase, both the number of privileged documents and precision tend to rise. However, with additional iterations, more entities receive non-zero scores, expanding the coverage of privileged documents—though this broader inclusion leads to a decline in precision.

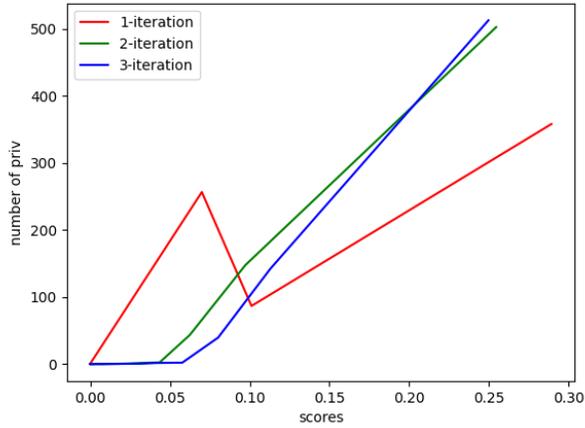

**Fig. 1.** The correlations between entity scores and the number of privileged documents of the first dataset.

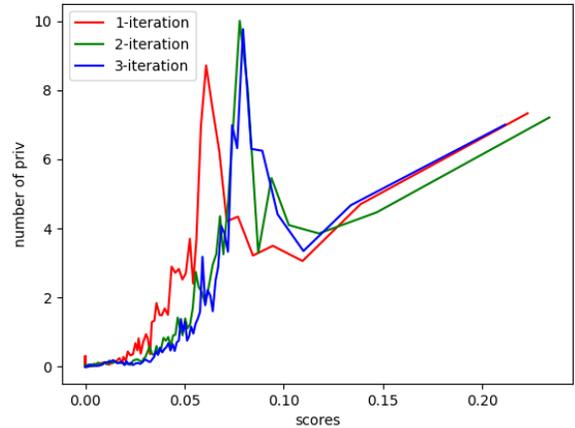

**Fig. 3.** The correlations between link scores and the number of privileged documents of the first dataset.

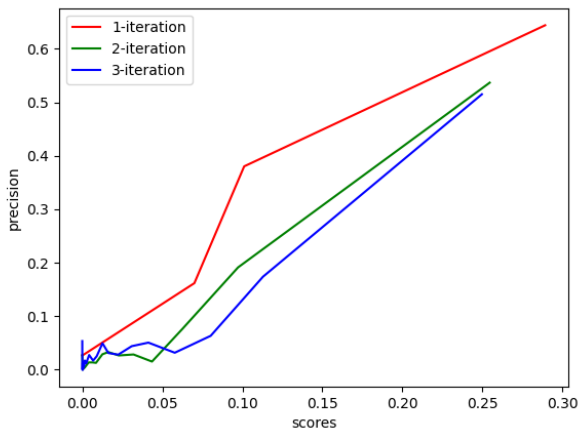

**Fig. 2.** The correlations between entity scores and precision of the first dataset.

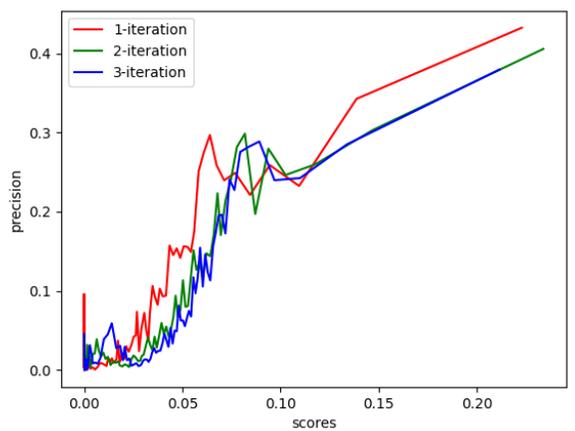

**Fig. 4.** The correlations between link scores and precision of the first dataset.

Figures 3 & 4 present the correlations between link scores and both the number of privileged documents and precision of the first dataset. These metrics—score, document count, and precision—are averaged across groups of 1,000 links. The figure displays curves corresponding to one, two, and three iterations of the algorithm. In general, the number of privileged documents increases with higher link scores, with a peak in the middle of the score range between 0.05 and 0.1. Precision also tends to rise with increasing scores. The results are similar across all three iterations.

Figures 5 & 6 illustrate the correlations between entity scores and both the number of privileged documents and precision for the second dataset. As with the first dataset, these metrics—score, document count, and precision—were averaged across groups of 1,000 entities. The results closely mirror those of the first dataset: as scores increase, both the number of privileged documents and precision generally rise. Differences across iterations are minimal.

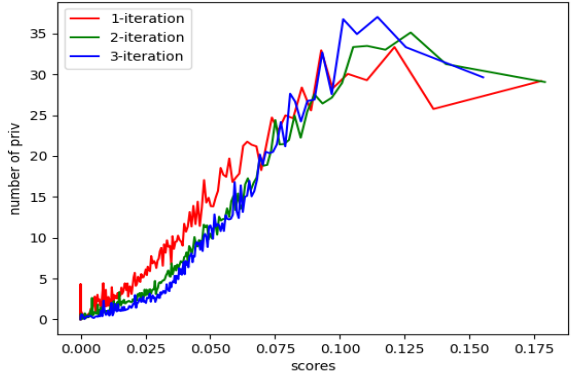

**Fig. 5.** The correlations between entity scores and the number of privileged documents of the second dataset.

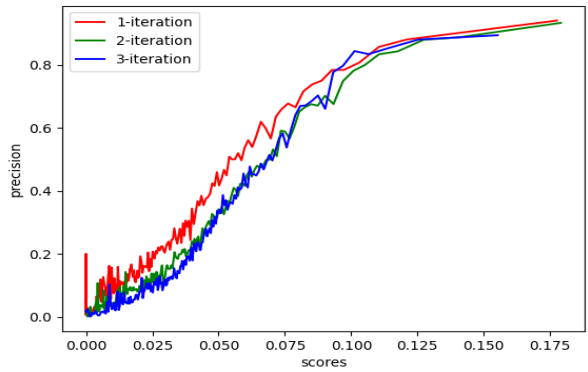

**Fig. 6.** The correlations between entity scores and precision of the second dataset.

Figures 7 & 8 present the correlations between link scores and both the number of privileged documents and precision in the second dataset. As link scores increase, both the number of privileged documents and precision generally rise.

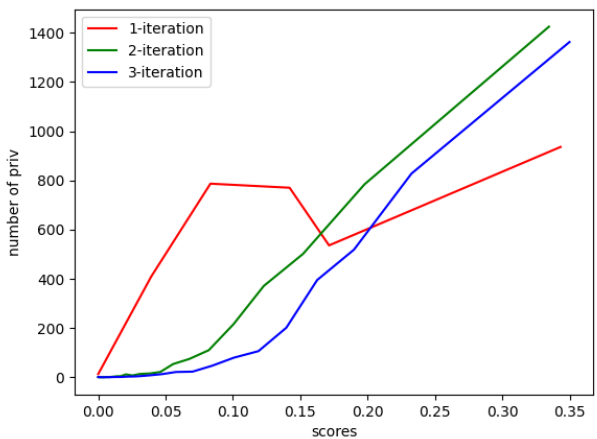

**Fig. 7.** The correlations between link scores and the number of privileged documents of the second dataset.

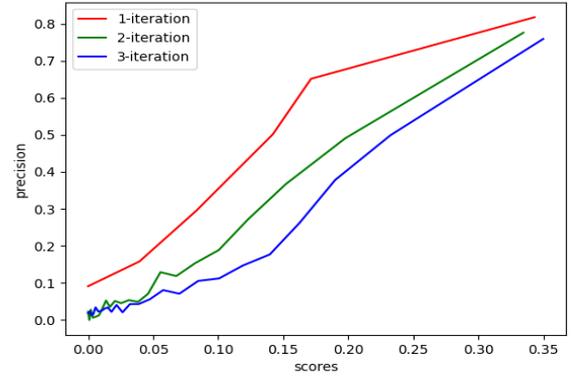

**Fig. 8.** The correlations between link scores and the precision of the second dataset.

Figures 9 and 10 depict the correlations between entity scores and both the number of privileged documents and precision within the third dataset. The results closely resemble those observed in the first two datasets: as scores increase, both the number of privileged documents and precision tend to rise. However, the variations across iterations remain substantial.

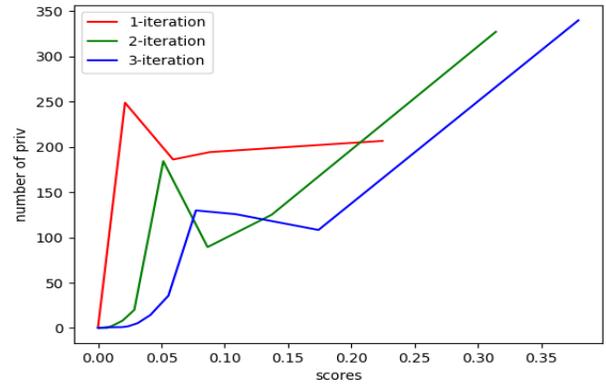

**Fig. 9.** The correlations between entitu scores and the number of privileged documents of the third dataset.

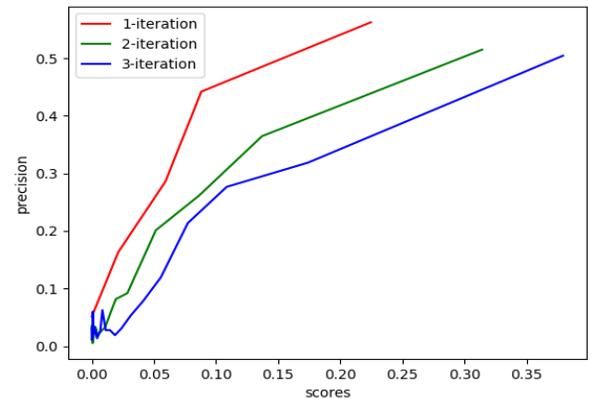

**Fig. 10.** The correlations between entity scores and precision of the third dataset.

Figures 11 and 12 illustrate the correlations between link scores and both the number of privileged documents and precision within the third dataset. As link scores increase, there is a general upward trend in both privileged document count and precision.

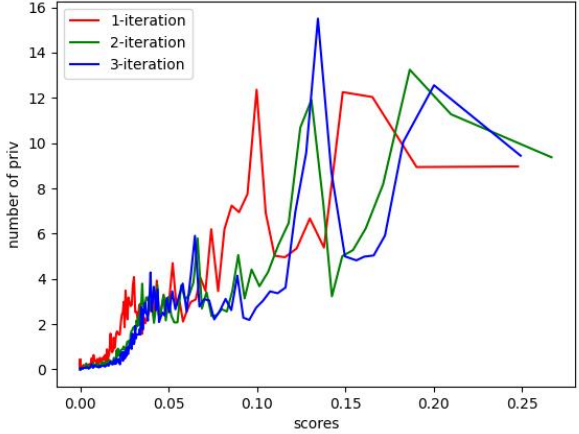

**Fig. 11.** The correlations between link scores and the number of privileged documents of the third dataset.

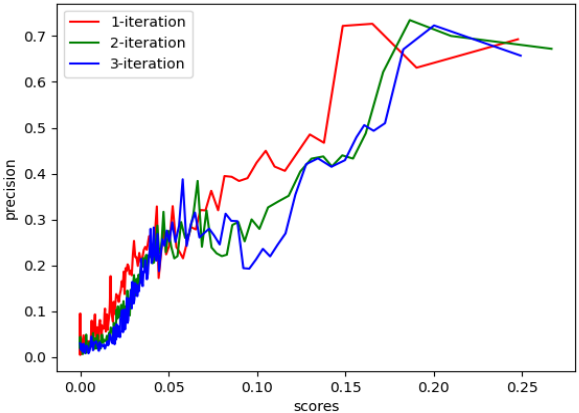

**Fig. 12.** The correlations between link scores and the number of precision of the third dataset.

Overall, the results indicate that the legal entity ranking algorithm performs effectively. High-scoring entities tend to be associated with a greater number of privileged documents and achieve higher precision. We anticipate that these ranked entities can assist in the identification of privileged content. With additional iterations, more entities acquire non-zero scores, broadening the coverage of privileged documents. This expanded inclusion, however, results in a decrease in precision.

### B. Second Experiment

The legal entity ranking algorithm is capable of assigning non-zero scores to entities not classified as counsel. The second set of experiments evaluates the effectiveness of non-counsel entities with scores of 0.1 or higher in identifying privileged documents. These experiments utilized the same three datasets as in the previous experiments. These entities were evenly divided into three groups: LikelyPriv1, LikelyPriv2, and LikelyPriv 3. All remaining entities were categorized as LikelyNonPriv. LikelyPriv1 comprises entities with the highest scores, LikelyPriv2 includes those with mid-range scores, and LikelyPriv3 consists of entities with the lowest scores.

In these experiments, we began by establishing links between entities, which were then categorized into groups such as LikelyPriv$N$. LikelyPriv$M$ (where N, M $\in$ {1, 2, 3}), LikelyPriv$N$.LikelyNonPriv, LikelyNonPriv.LikelyPriv$N$, and LikelyNonPriv.LikelyNonPriv. A link is classified as LikelyPriv$N$.LikelyPriv$M$ if the email sender belongs to LikelyPriv$N$ and the receiver to LikelyPriv$M$. Recall and precision were calculated for each link category. A document is classified as privileged by a link category if a link from this category occurs in the document. We hypothesize that Relevant1.Relevant1 will exhibit the highest precision, while LikelyNonPriv.LikelyNonPriv will show the lowest, given that entities in *Relevant1* have the highest scores and those in Other have scores below 0.1. Tables II, III, and IV present the recall and precision metrics for these categories across the three datasets, respectively.

In all three datasets, the LikelyNonPriv.LikelyNonPriv category consistently yielded the lowest precision rates, falling below the baseline of random guessing (0.28 for the first dataset, 0.165 for the second, and 0.244 for the third). The precision rate for Relevant1.Relevant1 ranks third highest in the first dataset, highest in the second, and the fourth in the third. Across all three datasets, the highest precision values consistently involve Relevant1. Overall, precision rates tend to decrease as entity scores decline, with lower-score categories generally exhibiting lower precision than their higher-score counterparts.

TABLE II.    RECALL AND PRECISION RATES FOR THE FIRST DATASET.

| Link Category | Recall | Precision |
|---|---|---|
| LikelyPriv1. LikelyPriv1 | 6.3% | 40.1% |
| LikelyPriv1. LikelyPriv2 | 4.2% | 39% |
| LikelyPriv1. LikelyPriv3 | 4.1% | 39% |
| LikelyPriv1. LikelyNonPriv | 6% | 41.1% |
| LikelyPriv2. LikelyPriv1 | 3.5% | 41.7% |
| LikelyPriv2. LikelyPriv2 | 3.9% | 23.4% |
| LikelyPriv2. LikelyPriv3 | 7% | 28.4% |
| LikelyPriv2. LikelyNonPriv | 7.5% | 23.8% |
| LikelyPriv3. LikelyPriv1 | 2.8% | 32.3% |
| LikelyPriv3. LikelyPriv2 | 7.4% | 32.1% |
| LikelyPriv3. LikelyPriv3 | 16.7% | 37.9% |
| LikelyPriv3. LikelyNonPriv | 20.3% | 31.6% |
| LikelyNonPriv. LikelyPriv1 | 4.8% | 37% |
| LikelyNonPriv. LikelyPriv2 | 10.4% | 19.6% |
| LikelyNonPriv. LikelyPriv3 | 23% | 21.1% |
| LikelyNonPriv. LikelyNonPriv | 25.2% | 12.5% |

TABLE III. RECALL AND PRECISION RATES FOR THE SECOND DATASET

| Link Category | Recall | Precision |
|---|---|---|
| LikelyPriv1. LikelyPriv1 | 26.8% | 55.6% |
| LikelyPriv1. LikelyPriv2 | 22.1% | 50.7% |
| LikelyPriv1. LikelyPriv3 | 17.5% | 49.5% |
| LikelyPriv1. LikelyNonPriv | 33.1% | 44.9% |
| LikelyPriv2. LikelyPriv1 | 17.9% | 46.1% |
| LikelyPriv2. LikelyPriv2 | 14.4% | 35% |
| LikelyPriv2. LikelyPriv3 | 12.9% | 33.6% |
| LikelyPriv2. LikelyNonPriv | 22.5% | 29.6% |
| LikelyPriv3. LikelyPriv1 | 11.9% | 41% |
| LikelyPriv3. LikelyPriv2 | 10.7% | 31.3% |
| LikelyPriv3. LikelyPriv3 | 8.6% | 29.1% |
| LikelyPriv3. LikelyNonPriv | 16.5% | 25.8% |
| LikelyNonPriv. LikelyPriv1 | 43.6% | 34.8% |
| LikelyNonPriv. LikelyPriv2 | 36.4% | 25.2% |
| LikelyNonPriv. LikelyPriv3 | 31.6% | 22.7% |
| LikelyNonPriv. LikelyNonPriv | 66.2% | 13% |

TABLE IV. RECALL AND PRECISION RATES FOR THE THIRD DATASET.

| Link Category | Recall | Precision |
|---|---|---|
| LikelyPriv1. LikelyPriv1 | 9.3% | 74.3% |
| LikelyPriv1. LikelyPriv2 | 10.4% | 75.8% |
| LikelyPriv1. LikelyPriv3 | 9.7% | 79% |
| LikelyPriv1. LikelyNonPriv | 4.7% | 59.6% |
| LikelyPriv2. LikelyPriv1 | 10.3% | 74.9% |
| LikelyPriv2. LikelyPriv2 | 15.8% | 49.5% |
| LikelyPriv2. LikelyPriv3 | 12.9% | 45.4% |
| LikelyPriv2. LikelyNonPriv | 10.1% | 38.6% |
| LikelyPriv3. LikelyPriv1 | 6.8% | 76.3% |
| LikelyPriv3. LikelyPriv2 | 9.9% | 39.5% |
| LikelyPriv3. LikelyPriv3 | 9% | 35.7% |
| LikelyPriv3. LikelyNonPriv | 8.1% | 28.6% |
| LikelyNonPriv. LikelyPriv1 | 2.9% | 45.1% |
| LikelyNonPriv. LikelyPriv2 | 9.6% | 28.9% |
| LikelyNonPriv. LikelyPriv3 | 8.8% | 20.8% |
| LikelyNonPriv. LikelyNonPriv | 62.4% | 18% |

## V. CONCLUSION AND FUTURE WORK

In conclusion, the proposed link analysis method demonstrates a promising approach to identifying privileged documents by leveraging the structure of human entity networks derived from email metadata. By assigning scores to entities based on their proximity to known lawyers and analyzing the strength of their connections, the algorithm effectively distinguishes individuals likely to be involved in privileged communications. The experimental results validate the utility of both entity and link scores in enhancing detection accuracy, offering a scalable and data-driven solution for legal document review. In privileged document review, we identified emails exchanged between two non-lawyers who were acting on behalf of a company lawyer. Their involvement was inferred from their strong professional ties to the legal team. This communication would not surface through a standard search for lawyer names, making this connection the only way to uncover it.

In our experiments, we observed that certain entities engaged in frequent communications with multiple lawyers, yet received low scores. This was primarily due to their extensive interactions with non-lawyers, which diluted their ranking. However, we believe these entities are highly likely to be involved in privileged communications and can play a key role in identifying privileged documents. To better reflect this, we plan to enhance our legal entity ranking algorithm to assign higher scores to such entities.

In the future, we can leverage information from email content to assign weights to links between entities. Specifically, privileged keywords found within the content can be used to influence these weights. Links containing strong privileged keywords should be assigned higher weights than those without. These link weights will then contribute to the calculation of entity scores, where higher link weights lead to higher entity scores.

When labeled privileged documents are available, they can be used to calculate link weights between entities. Links associated with privileged documents should be assigned higher weights, while those without such associations should receive lower weights.

## REFERENCES

[1] S. Brin and L. Page. The anatomy of a large-scale hypertextual Web search engine. Computer Networks and ISDN Systems, 30(1–7):107–117, 1998.

[2] L. Page, S. Brin, R. Motwani, and T. Winograd. The pagerank citation ranking: Bringing order to the web. Technical report, Stanford Digital Libraries SIDL-WP-1999-0120, 1999.

[3] J. M. Kleinberg. Authoritative sources in a hyperlinked environment. Journal of the ACM, 46(5):604–632, September 1999.

[4] Prithviraj Sen, Galileo Namata, Mustafa Bilgic, Lise Getoor, Brian Galligher, and Tina Eliassi-Rad. Collective classification in network data. AI magazine, 29(3):93–93, 2008.